\documentclass[a4paper,conference,final]{IEEEtran}

\usepackage{siunitx}
\DeclareSIUnit \dBm {dBm}
\DeclareSIUnit \dB {dB} 
\DeclareSIUnit \dBi {dBi} 
\DeclareSIUnit \Kbps {Kbps}
\DeclareSIUnit \Mbps {Mbps}
\DeclareSIUnit \Gbps {Gbps}
\DeclareSIUnit \kBps {kBps}
\DeclareSIUnit \MBps {MBps}
\DeclareSIUnit \GBps {GBps}

\usepackage{cite}      
\usepackage[caption=false, font=footnotesize]{subfig}
\usepackage[dvips]{graphicx}
\usepackage{url}       
\usepackage{amsfonts}
\usepackage{amssymb}
\usepackage{mathtools}
\usepackage{times}
\usepackage{algpseudocode}
\usepackage{fourier}
\usepackage{algorithm}
\usepackage{enumerate}
\usepackage{multirow}
\usepackage{tikz}
\usepackage{array}
\usepackage{gensymb}
\usepackage{eucal}
\newcolumntype{P}[1]{>{\centering\arraybackslash}p{#1}}
\newcolumntype{M}[1]{>{\centering\arraybackslash}m{#1}}

\DeclareMathOperator*{\maximise}{\mathrm{Maximise}}

\bstctlcite{IEEEexample:BSTcontrol}
\newcommand*{\maximisel}{\maximise\limits}

\algnewcommand\algorithmicinputttt{\textbf{Input:}}
\algnewcommand\Input{\item[\algorithmicinputttt]}

\algnewcommand\algorithmicinput{\textbf{Output:}}
\algnewcommand\Output{\item[\algorithmicinput]}

\algnewcommand\algorithmicinputt{\textbf{Phase 1:}}
\algnewcommand\Phase{\item[\algorithmicinputt]}

\algnewcommand\algorithmicinputtt{\textbf{Phase 2:}}
\algnewcommand\Phasee{\item[\algorithmicinputtt]}

\usepackage[english]{babel}
\usepackage{blindtext}

\begin{document}
\title{Efficient V2V Communication Scheme for 5G MmWave Hyper-Connected CAVs}
\author{\IEEEauthorblockN{Ioannis Mavromatis, Andrea Tassi, Robert J. Piechocki and Andrew Nix}
  \IEEEauthorblockA{Department of Electrical and Electronic Engineering, University of Bristol, UK \\ Emails: \{ioan.mavromatis, a.tassi, r.j.piechocki, andy.nix\}@bristol.ac.uk}
}

\maketitle

\begin{abstract}
Connected and Autonomous Vehicles (CAVs) require continuous access to sensory data to perform complex high-speed maneuvers and advanced trajectory planning. High priority CAVs are particularly reliant on extended perception horizon facilitated by sensory data exchange between CAVs. Existing technologies such as the Dedicated Short Range Communications (DSRC) are ill-equipped to provide advanced cooperative perception service. This creates the need for more sophisticated technologies such as the 5G Millimetre-Waves (mmWaves). In this work, we propose a distributed Vehicle-to-Vehicle (V2V) mmWaves association scheme operating in a heterogeneous manner. Our system utilises the information exchanged within the DSRC frequency band to bootstrap the best CAV pairs formation. Using a Stable Fixtures Matching Game, we form V2V multipoint-to-multipoint links. Compared to more traditional point-to-point links, our system provides almost twice as much sensory data exchange capacity for high priority CAVs while doubling the mmWaves channel utilisation for all the vehicles in the network.
\end{abstract}


\vspace{-2mm}
\section{Introduction}

Nowadays, driverless vehicles are finding increased applicability for commercial usage and emergency services. The trajectory of a vehicle not executing emergency maneuvers (e.g., avoiding a pedestrian),
tends to follow a well-defined path~\cite{trajectory_planning}. The short-term trajectory planning can be achieved by fusing its onboard sensor information, acquired by LIDARs, camcoders, etc.
However, the long-term planning is based on the ``knowledge'' acquired from the surrounding vehicles~\cite{connected_path_planning}. 
Hence, the exchange of sensor data becomes mandatory and a reliable
\emph{Internet of CAVs} is of paramount importance. A 5G communication framework will be able to deliver reliably and rapidly~\cite{7858628} the crucial and demanding sensor data maximising the road safety~\cite{2018arXiv180109510M,8275608}.

Connected Autonomous Vehicles (CAVs) are estimated to generate up to \SI{1}{\Gbps} of sensor data. In order to achieve cooperative maneuvering on a road, these data should be exchanged subject to strict Quality-of-Service (QoS) constraints~\cite{sensors}.
Millimetre Waves (mmWaves) can fulfill these requirements, having as a drawback the increased signal attenuation, especially under greater distances. Within urban areas though, where CAVs are in close proximity, IEEE 802.11ad mmWaves protocol can achieve up to \SI{6.75}{\Gbps} of data rate under Line-of-Sight (LOS) conditions~\cite{TassiTVT}.

Comparing this value with the previous estimation regarding the sensor data, it is obvious that forming point-to-point links will lead to under-utilisation of the channel. Also, \emph{emergency CAVs} (e-CAVs), compared to \emph{regular CAVs} (r-CAVs), due to their increased size and speed introduce more driving risks while performing obstacle avoidance maneuvers and advanced trajectory planning~\cite{emergency_vehicles}. To that extent, they require prioritized connectivity, without though dictating the wireless channel as this may lead to limited access to sensor data for the remaining r-CAVs. 



To overcome the above problems, we will present a multi-link association scheme for mmWave Vehicle-to-Vehicle (V2V) communications, able to better utilise the channel and prioritize the exchange of sensor data for the e-CAVs without limiting the acquired information of the r-CAVs. The adoption of mmWaves for point-to-point vehicular communications is receiving significant attention in the literature, e.g.~\cite{beam_design},~\cite{mywork2}. 
Forming multiple links with the nearby CAVs can significantly enhance the performance of a system. Therefore, in our work we will formulate a sophisticated many-to-many link association scheme able to maximise the channel utilisation by taking into account the achievable data rate of mmWave channels, the different types of CAVs and their movement and positioning.

We will formulate our association policy as a distributed constraint optimisation problem, where the information required will be exchanged via Dedicated Short Range Communication (DSRC) links in a heterogeneous manner. To solve our problem, we will use the Matching Theory framework~\cite{matching_theory}.
A similar approach can be found in~\cite{wysiwyg}, where authors used Gale-Shapley's Deferred Acceptance matching algorithm. The CAVs were grouped into transmitters and receivers forming many-to-one V2V links, limiting  though the system in this unidirectional relationship (many TXs linked to one RX). In our work, using the Stable Fixtures (SF) matching game~\cite{stable_fixtures}, we will formulate a more generalized many-to-many link association scheme forming bidirectional links and taking into account the unique characteristics of each CAV.

This paper is organized as follows. In Sec.~\ref{sec:system_model} we present our system model. We start with our link budget analysis and introduce our matching game and the concept of the matching capacity. Sec.~\ref{sec:matching_game} describes our problem formulation, the V2V SF matching game and the definitions and constraints for a stable matching. Sec.~\ref{sec:perf_evaluation} describes our simulation setup and presents our performance investigation. This paper is concluded in Sec.~\ref{sec:conclusions} summarising our finding.

\vspace{-1mm}
\section{System Model}\label{sec:system_model}

Consider an urban vehicular scenario where mmWaves links are established between moving vehicles in a V2V manner. Let $\mathcal{V} \triangleq \left \{1,\ldots,V \right \}$ denote the number of CAVs in the system. CAVs are of two different types: e-CAVs, denoted as $\mathcal{E} \triangleq \left \{1,\ldots,E \right \}$ and r-CAVs, denoted as $\mathcal{R} \triangleq \left \{1,\ldots,R \right \}$ with $\mathcal{E},\mathcal{R}\subseteq\mathcal{V}$, $\mathcal{E}\cup\mathcal{R}=\mathcal{V}$, and $\mathcal{E}\cap\mathcal{R}=\emptyset$. 

Links are half-duplex, interconnecting two CAVs for a timeslot with duration $T_{s}$. Let $A_{j}$ be the number of CAVs linked with CAV $i$ within a $T_{s}$. The formed links are the pairs $\sum\nolimits_{i \in \mathcal{V}} \left[ \left(i,j\right):j \in A_{j} \right]$. We assume that a scheduling algorithm shares the available time between all the pairs in equal transmission slots $\mathcal{T}$ with a duration of $\left\lbrace\mathcal{T}_{t}: T_{s}\mod\mathcal{T}_{t}=0 \right\rbrace$.

\vspace{-1mm}
\subsection{MmWave Link Budget analysis}\label{sub:link_budget}

The IEEE 802.11ad standard~\cite{standard} defines a sensitivity threshold $K_{MCS}$ for each Modulation and Coding Scheme (MCS). For a link $(i,j)$ between two vehicles at time $t \in \mathcal{T}_{t}$ and a given Signal-to-Interference-plus-Noise Ratio (SINR), an appropriate MCS can be chosen based on $SINR_{i,j} \geq K_{MCS}$. Knowing the chosen MCS, we know the maximum achievable data rate $r^{\mathrm{max}}_{i,j}(t)$ for this link.

The $\mathrm{SINR}_{i,j}$ is time dependent and defined as the ratio between the \emph{received power} $ P_{i,j}^\mathrm{rx}$  over the \emph{noise power} $P_{\mathrm{n}}$ plus the \emph{interference} $\mathcal{I}(t)=\sum\nolimits_{k \in \mathcal{V}, k \neq j} I_{k,j}(t)$, i.e., $\mathrm{SINR}_{i,j}(t) = P_{i,j}^\mathrm{rx}(t)/(P_{\mathrm{n}} + \mathcal{I}(t))$. 
$P_{i,j}^\mathrm{rx}$ is given as~\cite{prediction_model}:
\begin{equation}
P_{i,j}^\mathrm{rx}(t) = P^\mathrm{tx} + G_{i,j}^\mathrm{rx}(t) + G_{i,j}^\mathrm{tx}(t) - PL_{i,j}(t)
\end{equation}
where $P^\mathrm{tx}$ is the transmission power, $G_{i,j}^\mathrm{tx}$ and $G_{i,j}^\mathrm{rx}$ are the TX and RX antenna gains, and $PL$ is the \emph{path-loss component}.

The ideal antenna beams are modelled with gain $G_{i,j}^\alpha$, with $\alpha \in \left\{ tx,rx \right\}$, and zero sidelobes. $G_{i,j}^\alpha$ is a function of the half-power beamwidth $\theta$ (\SI{-3}{\dB}), is equal for both polarisation planes and is given as~\cite{balanis}:
\begin{equation}
G_{i,j}^\alpha(t) \simeq 4 \, \pi/\theta_{i,j}^2(t),~~\mathrm{for}~ \left| \vartheta_{i,j}(t) \right| \leq \theta_{i,j}(t)/2
\end{equation}
where $\vartheta_{i,j}(t)$ is the misplacement error that leads to misalignments. In this work, we assumed perfect beam alignment all the time. The $PL$ is calculated as:
\begin{equation}
PL_{i,j}(t) = 10 \, n \, \log_{10}d_{i,j}(t) + C_{i,j}^\mathrm{att}(t) + S_{\mathrm{f}}(t) 
\end{equation}
where $n$ is the path-loss exponent and $d_{i,j}$ is the distance between vehicles $i$ and $j$. $C_{i,j}^\mathrm{att}$ is the channel attenuation with regard to the distance $d_{i,j}$ at time $t$, given by the rain and atmospheric attenuation and the channel attenuation factor $H^\mathrm{att}$ for a mmWave LOS link at \SI{60}{\giga\hertz} in urban environments~\cite{prediction_model}, i.e. $C_{i,j}^\mathrm{att}(t) = 40d_{i,j}(t)/1000 + H^\mathrm{att}$. Finally, $S_{\mathrm{f}}$ represents the shadow fading of the channel following a log-Normal distribution $S_{\mathrm{f}}\sim\log\mathcal{N} (0,\sigma^2_{S_f})$ with $\sigma = 5.8$~\cite{sigma_sf}. Term $P_{\mathrm{n}}$ can be calculated as: $P_{\mathrm{n}} = N_{\mathrm{fl}}+10\log_{10}B+N_{\mathrm{fg}}$, with $N_{\mathrm{fl}}$ being the noise floor value and $N_{\mathrm{fg}}$ the noise figure.

With remark to the aforementioned link budget analysis, the average data rate of the vehicle $i$ for all the links $\sum\nolimits_{i \in \mathcal{V}} \left[ \left(i,j\right):j \in A_{j} \right]$ is given as:
\begin{equation}
C_{i}(t) = \frac{1}{\mathcal{T}} \displaystyle \sum\limits_{j \in A_{j}} r^{\mathrm{max}}_{i,j}(t)
\end{equation}

\begin{figure}[t]     
\centering
    \includegraphics[width=0.95\columnwidth]{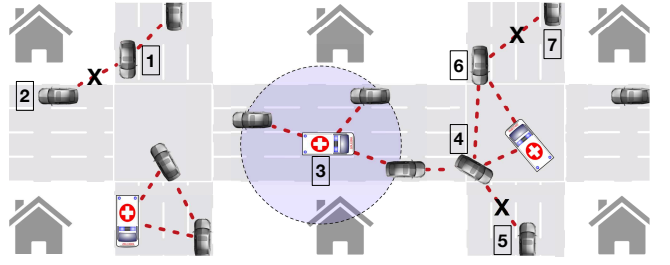}
    \vspace{-2mm}
    \caption{A snapshot of our system at time $t$, generated from the SF matching game using all the utility functions $\mathcal{U}$. Links are formed  (dotted lines) or not (lines crossed by an ``X'') according to the maximisation problem introduced. The capacity was assumed to be equal to 3.}
    \label{fig:v2vSystem}
\end{figure}

\vspace{-2mm}
\subsection{Matching Capacity for the Association Policy}\label{sub:capacity}
\vspace{-1mm}
In this system a CAV $i$ can be potentially linked with all CAVs at its $360\degree$~azimuth plain under some specific conditions that will be described later. At time $t$, $i$ is associated with $A_j$ vehicles, of which $A_{e}$ are e-CAVs and $A_{r}$ are r-CAVs with $A_{r},A_{e} \in \mathbb{N}^*, A_{r},A_{e} \subseteq \mathcal{V}$ and $A_j = A_{r} + A_{e}$.

$A_j$ can be affected by the number of vehicles in LOS. In fact, the high radio frequency of mmWaves significantly affects the received signal strength and even small obstacles (e.g. another CAV) can behave as impenetrable blockages. To that extent, we formulate our problem so that potential CAVs for association are always in LOS with our target $i$. For example, in Fig.~\ref{fig:v2vSystem} CAV no. $1$ cannot be paired with CAV no. $2$, as their link is blocked by a building.


The matching capacity $c_i \in \mathbb{N}^*$ is defined as the maximum number of CAVs that $i$ can be associated with. The capacity is limited by two factors. Firstly, the maximum number of links that each CAV can establish (predefined at the beginning of each scenario). Secondly, is capped with regard to the exchanged data within a timeslot, i.e. $C_i(t) \leq r^{max}_{i,j}(t)$ should always be true. CAV no. $4$, for example cannot be linked with with CAV no. $5$ (Fig.~\ref{fig:v2vSystem}) as its matching capacity was reached (connected to three CAVs with higher priority).

\vspace{-1mm}
\subsection{Requirements for the System Evaluation}
\vspace{-1mm}
A cooperative maneuvering on a road can be achieved by sharing the generated sensor data with the surrounding CAVs. Different amounts can be generated per CAV at different timeslots. This depends on the resolution and the number of equipped sensors, the surrounding environment and the preprocessing that might take place before the transmission. This amount can be quantified as $g \in \left\{ 1,\ldots,G \right\}$, where $g$ is measured in data transfer rate units.

The main idea behind our systems is that a CAV, evaluating the importance and utility of a close vehicle can decide whether it wants to connect with that or not. This potentially formed link $(i,j)$, can be quantified with a utility function $u_{i,j}(\cdot)\in [0,1]$ and is based on the opposing vehicles topologies and movement, the generated sensor information, and finally, their types of vehicle. 

Our system operates in a heterogeneous manner (DSRC/mmWaves) and the information required for the utility function are encapsulated within the DSRC beacon messages (as in~\cite{mywork2}).
The generated data $g$ are estimated from each CAV for the next scheduling period and accompanied with the aforementioned information, they are being broadcast in the network.
Using this heterogeneous architecture, we leverage from the increased coverage of the DSRC ensuring that vehicles with close proximity will receive the same information. Also, the reduced signal loss minimises the error rate in the beacon reception at close distances. When the information from the surrounding vehicles is received, all CAVs can independently run their SF matching game. The utilities for all CAVs are represented as $\mathcal{U} = \left\lbrace u_{i,j} | i,j \in \mathcal{V} \right\rbrace$. The variables constituting the above utility function are the following:

\subsubsection{Type of Vehicle}
The type of vehicle, i.e. $\tau_{i} \in \left\lbrace \mathrm{e-CAV}_{i},\mathrm{r-CAV}_{i}: \mathrm{e-CAV}_{i}=1,\mathrm{r-CAV}_{i}=0.5 \right\rbrace$, is a very important factor. The traffic of a CAV can be prioritized towards an emergency vehicle as the necessity for more complex maneuvers requires access to more sensor data.

\subsubsection{Regional Sensor Data Generated}\label{sub:regional}
The density $q_{i}$ can be characterized as the number of CAVs $q \in \left\{ 1,\ldots,K \right\}$  with $K \subseteq \mathcal{V}$ within a circular radius $R \in \mathbb{N}^*$. $q_{i}$ is affected by the urban topology (e.g., buildings cover a large surface area that vehicles cannot use) and the road characteristics (e.g., more road lanes imply that more CAVs can be positioned within $R$). In Fig.~\ref{fig:v2vSystem}, the dotted circle centered around CAV no. $3$, represents its sensor region $R$. Targeting a section of a road with increased density implies that we can achieve a better link utilisation and access to more sensor data, especially necessary for e-CAVs. What is more, the shorter distance to the target CAVs will enhance the perceived SINR.
The regional data generated for CAV $i$ can be given as $Q_{i}=g_{i}+\sum\nolimits_{\substack{j=1,\ldots,q_i}}g_{j}$ and this value can be normalized with respect to the maximum regional data generated per timeslot, i.e. $Q^*_{i}(t)=\frac{Q_{i}(t)}{\max (Q(t))} \in \left[ 0,1 \right]$.

\subsubsection{Vehicle Direction and Distance}
We assume that a CAV is more interested in data coming from CAVs that their routes coincide with the route of the later. This is important because vehicles can collect more sensor information for an unknown road section. However, the utility of the information degrades with respect to the distance. To that extent, in order to take into account both variables we use the normalized weighted arithmetic mean as below:
\begin{equation}
e_{i,j} = w_1\frac{R-d_{i,j}}{R} + w_2\frac{\pi - \phi^{i,j}_n}{\pi}~~~\in \left[ 0, 1 \right]
\end{equation}
where $\phi^{i,j}_n$ is the normalized angle difference between the two vehicle directions and $w_1$ and $w_2$ are the weight factors. Values $\approx 1$ determine a coinciding movement and very close distance of the vehicles.

For all the above quantifying parameters, a utility function can be written as $u_{i,j}(\left\lbrace \tau_{j}, Q_{j}, e_{i,j} \right\rbrace) \in \left[ 0,1 \right]$, evaluating the link $(i,j)$ between two vehicles. More specifically, the more valuable a CAV is, the more important is the connection with that. The utility function is given as:
\begin{equation}
u_{i,j}(t) = \tau_{j} \, Q_{j}(t) \, e_{i,j}(t) \, \frac{r_{i,j}(t)}{r^{\mathrm{max}}_{i,j}(t)}~~\in \left[ 0,1 \right]
\end{equation}

\vspace{-7mm}
\section{Matching Game and Problem Formulation}\label{sec:matching_game}
\subsection{Problem Definition}\label{sub:problemdefinition}
The communication links that can potentially be established between all vehicles at time $t \in T_{s}$ are defined with a matrix $\mathcal{M}(t) = \left( m_{i,j}  \right) \in \mathbb{R}^{x \times y}$, $x,y \leq \mathcal{V}$ where $m_{i,j}$ is a binary value representing the existence of a link between vehicles $i$ and $j$, having the following values:
\begin{equation}\label{eq:matching}
m_{i,j}(t) = 
  \begin{cases}
      1,~\mathrm{link}~(i,j)~\mathrm{is~established~for}~\left(t,t+\mathcal{T}_{t}\right]\\
     0,~\mathrm{otherwise}
  \end{cases}
\end{equation}

Each CAV can be simultaneously matched with less or equal than $c_{j} \in \mathbb{N}^*$ vehicles, being either e-CAVs or r-CAVs. This value depends on the matching capacity of each vehicle as described in Sec.~\ref{sub:capacity} and~\ref{sub:regional}. Let $\mathcal{V}_{j} \subseteq \mathcal{V}$ be the number of CAVs paired with $i$ at time $t$. The previously introduced utility function should be redefined as:
\begin{equation}\label{eq:redefined_fitness}
\mathcal{X}_i(\mathcal{M}(t)) \equiv  \mathcal{X}_i \left( \left\lbrace \tau_{j}, Q_{j}, e_{i,j} \right\rbrace_{j \in \mathcal{V}_{j}(t)} \right)~~\in \left[ 0,1 \right]
\end{equation}
representing the utility functions of all the vehicles that can be potentially paired with $i$.
The problem can be formulated as the maximisation of \eqref{eq:redefined_fitness} in order to find the best links between all CAVs at time $t$:
\begin{subequations}\label{eq:problem}
\begin{align}
\displaystyle \maximisel_{\mathcal{M}(t)} & \displaystyle\sum\limits_{i \in \mathcal{V}_{i}(t)}\mathcal{X}_i(\mathcal{M}(t)) & \\ 
\textrm{subject to:} & \displaystyle \sum\limits_{j \in \mathcal{V}_{j}(t)} m_{i,j}(t) \leq c_{j} & , \forall j \in \mathcal{V}(t) & \\
 & j~\mathrm{in~LOS~with}~i & , \forall j \in \mathcal{V}_{j}(t) & \\
 & j~\mathrm{within}~R  & , \forall j \in \mathcal{V}_{j}(t) & \\
 & m_{i,j}(t) \in \left( 0,1 \right] & , \forall i,j \in \mathcal{V}_{i}(t) \times \mathcal{V}_{j}(t) & \\
 & m_{i,j}(t) = 0 & , \forall i = j &
\end{align}
\end{subequations}
where the above constraints are the number of CAVs that can be associated with the target $i$ and the range of values for the utility function in \eqref{eq:redefined_fitness}.

\vspace{-1mm}
\subsection{Distributed Stable Fixtures (SF) Matching Game}

The problem formulated in \eqref{eq:problem} is a multigraph problem that is at least NP-complete. To solve it using a reasonable amount of resources we utilised a SF matching game with a complexity of $O(m)$~\cite{stable_fixtures}. Using this framework, we developed a strategy able to run independently on each CAV and establish the V2V links with the highest utility until the capacity is reached.
In the next sections, we will describe the basic definitions of the SF game with respect to our problem formulation in Sec.~\ref{sub:problemdefinition} and the two-phase operation required for the link establishment.


\vspace{-1mm}
\subsection{Definitions for the Distributed SF Matching Game}\label{sub:definitions}
A matching game is defined by a set of vehicles $\mathcal{V}(t) = \left\lbrace v_1, v_2, \ldots, v_n \right\rbrace$ where $n$ is the number of all CAVs. 
We denote $\sum\nolimits_{i,j \in \mathcal{V}(t)} \left(i,j\right)$ as the number of potential links that will be evaluated within this matching game. Also, we denote $m$ as a matching pair. We assume a strict preference relation for CAV $i$ between vehicles $\left( j, j' \right)$ with $j \neq j'$ and denoted it by $\succnsim_i$ that is complete, reflexive and transitive. For the two potential matches $\mathcal{M}(t)$ and $\mathcal{M}'(t)$ we have  $m_i(t) = j$ and $m'_i(t) = j'$. The above implies that:
\begin{equation}\label{eq:preference_relation}
\left( j, \mathcal{M}(t)  \right) \succnsim_i \left( j', \mathcal{M}'(t) \right) \iff u_{i,j}(t) > u_{i,j'}(t)
\end{equation}
meaning that $i$ prefers $j$ over $j'$ due to its higher utility. Intuitively, a stable matching refers to a matching where no players have the intention to change their preferences (switch preference for a higher prioritized CAV).

\textbf{\emph{Definition 1:}} \emph{A link $(i,j)$ is pairwise stable if and only if both $i$ and $j$ decide to associate with each other.}

\noindent This implies that there is a mutual approval from both vehicles and there are no \emph{blocking pairs}. For example, for CAVs $i,j,k \in \mathcal{V}(t)$ and $i \neq j \neq k$, we have $\left( u_{i,j}(t)>u_{i,k}(t) \right)$ and $\left( u_{j,i}(t)>u_{j,k}(t) \right), \forall k \in \mathcal{V}(t)$. Taking the comparison of the utilities and~\eqref{eq:preference_relation}, we have that $\succnsim_i$ and $\succnsim_j$ are complete and reflexive. Also, because this works for $\forall k \in \mathcal{V}(t)$, the preference relations are transitive as well.

\textbf{\emph{Definition 2:}} \emph{A link $(i,j)$ is group-wise stable if and only if it is not blocked by any coalition of vehicles $S \subseteq \mathcal{V}$.}

\noindent For a many-to-many matching game, the stability concept should focus not only on the blocking pairs between vehicles, but also on the blocking groups involved in the matching game.

If all the links are pairwise and group-wise stable at time $t$, we can ensure that SF will agree on a stable match. There is always a chance that a link does not meet these requirements. For example, in Fig.~\ref{fig:v2vSystem}, CAVs no. $6$ and $7$ formed a link at first, however, during the execution of the algorithm, it was discovered that CAV no. $6$ prefers to be linked with two other vehicles. In that case, the rejected link is removed from the matching game as it will be described in the next section.

\vspace{-1mm}
\subsection{Stability Criteria of SF Matching Game for V2V links}\label{sub:stability}

The general idea of the SF matching game is that each participating CAV starts with a fixed capacity (defined as the matching capacity $c_i$ before) and builds its preference list $P$ of potential CAVs to be matched with. We denote as $d_i = \min\left( c_i, |P_i| \right)$ the \emph{degree} of each CAV $i$ where $|P_i|$ is the length of its preference list. Each $P_i$ is classified as \emph{short} ($|P_i| < d_i$) or \emph{long} ($|P_i| > d_i$). SF executes in two phased, reducing the length of $P$ during each phase and concluding to the final matches. This is subject to two stability criteria: 

\subsubsection*{1}$\sum\nolimits_i d_i$ is not odd. This sum is double the size of a stable matching as it counts every match pair exactly twice, so if it is odd will imply that there is no pairwise and groupwise stability (as discussed in Sec.~\ref{sub:definitions}).

\subsubsection*{2}None of the preference lists $P_i$ becomes short during the execution of SF algorithm, i.e. $|P_i| \nless d_i, \forall i \in \mathcal{V}$. If any list is short, then this implies the elimination rotation was unsuccessful (more details about that will be given later) thus the algorithm reports the instance as unsolvable. 


\vspace{-1mm}
\subsection{The two Phases of the SF Matching Game}
Right now, we will describe the two phases of the SF matching algorithm. Algorithm~\ref{alg:sf_algorithm} presents these two phases with an pseudocode format as well.

\textbf{Phase 1:} During this phase, a sequence of \emph{bids} takes place from one vehicle to the others. These bids are used to construct a set $S$ that is an initial list of the potential matching pairs as well as to identify and delete pairs that cannot belong in a stable matching. For the pair $\left( v_i,v_j \right)$, $v_j$ is a \emph{target} for the $v_i$, and $v_i$ is a \emph{bidder} for the $v_j$. The target set for $v_i$ is denoted as $A_i = \left\lbrace v_j : \left( v_i,v_j \right) \in S \right\rbrace$ and the bidder set as $B_i = \left\lbrace v_j : \left( v_j,v_i \right) \in S \right\rbrace$. We also denote as $a_i = |A_i|$ and $b_i = |B_i|$ the length of each set. Each $v_i$, bids for its most favourable target and adds the pair $\left( v_j,v_i \right)$ in $S$ (line~\ref{line:line3}). Keep in mind that the pairs in $S$ are ordered so $\left( v_i,v_j \right)$ is different than $\left( v_j,v_i \right)$. Then $v_j$ checks whether it has exceeded its $B_j$ capacity, i.e. $b_j \geq c_j$ (line~\ref{line:line4}). If so, it deletes all vehicles within $P_j$ that are worse than the $c_j$th rank (lines~\ref{line:line5}-\ref{line:line11}). The bidding continues as long as $a_i < \min\left( c_i,|P_i| \right), \forall v_i \in \mathcal{V}$. The outcome of Phase 1 is a reduced $P_i$ for each $v_i$ and an increased set $S$ of potential matching pairs.

\textbf{Phase 2:} The two stability criteria defined in Sec.~\ref{sub:stability} are checked throughout the execution. If any becomes true, then the instance is reported as unsolvable (lines~\ref{line:line14} and~\ref{line:line20}). During this phase, we search for possible \emph{rotations} that can further reduce the length of $P_i$ lists. As a rotation is defined a sequence of ordered pairs $\rho = \left( \left( v_{i_0}, v_{j_0} \right), \left( v_{i_1}, v_{j_1} \right), \ldots, \left( v_{i_{r-1}}, v_{j_{r-1}} \right) \right)$. To find $\rho$, we start with a CAV having a long $P$. For each $k \in \left[ 0,r-1 \right]$ we denote $v_{i_k} = v_{l_(j_k)}$ and $v_{j_{k+1}} = v_{f_(i_k)}$. This means that $v_{i_k}$ is $v_{j_k}$'s worst bidder, $v_{j_{k+1}}$ is $v_{i_k}$'s next target and the pair $\left( v_{i_{k}},v_{j_{k}} \right)$ is in $\rho$. The process stops when a CAV has be visited twice. For each $\left( v_i,v_j \right)$ within $\rho$, this pair is removed (line~\ref{line:line19}) from $S$ and is replaced with the next favourable pair for $v_i$, found from its $P_i$ list (line~\ref{line:line23}). The rotation process stops when a preference list becomes short, i.e., there is no stable matching, or when no further possible rotations exist, i.e. a stable matching was found. More details about the SF matching algorithm such as the mathematical proofs and lemmas as well as the complexity analysis of the algorithm can be found in~\cite{stable_fixtures}.

\begin{algorithm}[t]

\caption{SF Matching Game Algorithm}
\label{alg:sf_algorithm}
{\footnotesize\begin{algorithmic}[1]
    \Phase
	\While{$a_i < \min\left( c_i,|P_i| \right)$} \Comment{$\forall i \in \mathcal{V}$}
    	\State Take $v_j$, the first vehicle in $P_i$, but not in $A_i$
    	\State $S = S \cup \left\lbrace ( v_i,v_j ) \right\rbrace$\label{line:line3} \Comment{Starting with $S$ as an empty set}
    	\If{$b_j \geq c_j$} \label{line:line4}
    		\State Find $v_k$ \label{line:line5}           \Comment $c_j\mathrm{th}$ ranked bidder for $v_j$ 
    		\Repeat~for each preferred vehicle $v_l$ over $v_k$ in $P_j$
            	\If{$( v_l,v_j ) \in S$}
                	\State $S = S \setminus \left\lbrace ( v_l,v_j ) \right\rbrace$.
                    \State Remove pair $( v_l,v_j )$ from $P$
                \EndIf
    		\Until{All $v_l$ removed from $P_j$} \label{line:line11}
    	\EndIf
    \EndWhile
    \Phasee
    \If{$\sum\nolimits_i d_i$ is odd}\label{line:line14}
		\State No stable matching exists
    \Else
    	\While{$|P_i| \nless d_i$}     \Comment{$\forall i \in \mathcal{V}$}
        	\State Find a possible rotation $\rho$ pair $( v_i,v_j )$  \Comment{Start with a long $P_i$}
            \State For $\rho$, remove $v_i$ and $v_j$ from preference lists \label{line:line19}
            \If{any $P_i$ becomes short} \label{line:line20}    \Comment{i.e. $|P_i| < d_i$}
            	\State No stable matching exists 
            \Else
            	\State $S = S \setminus \left\lbrace ( v_i,v_j ) \right\rbrace$ \label{line:line23}
            \EndIf
        \EndWhile
    	\State Return $\mathcal{M} = S$ \Comment{Stable Matching $\mathcal{M}$}
    \EndIf
\end{algorithmic}}
\end{algorithm}

\vspace{-1mm}
\section{Performance Evaluation}\label{sec:perf_evaluation}
\vspace{-1mm}
\subsection{Simulation Framework}

For our performance investigation we used a Manhattan Grid road network (\SI{100}{\meter}$\times$\SI{100}{\meter}), consisting of three horizontal and perpendicular roads. Each road is four-lanes wide (two lanes per direction) and each lane is \SI{3.2}{\meter}-wide. Using the SUMO traffic generator~\cite{sumo} we generated realistic vehicle traces for both the e-CAVs and r-CAVs. A vehicle can be an e-CAV with probability of $15\%$, implying that this vehicle will move with twice the speed of a r-CAV and will be impatient, i.e., will make maneuvers to overtake other vehicles and will be willing to impede other vehicles with higher priority on the road (e.g., crossing an intersection without using the right-of-way rule). For our channel model~\cite{prediction_model}, we used a carrier frequency of \SI{60}{\giga\hertz}, a bandwidth $B=\SI{2.16}{\giga\hertz}$, a path-loss exponent $n=2.66$~\cite{path_loss}, a transmission power $P^{tx}=\SI{10}{\dBm}$ and a channel attenuation $C^{att}=\SI{70}{\dBm}$~\cite{prediction_model}. For $P_n$, the noise figure $N_{fg}=\SI{6}{\dBm}$ and the noise floor $N_{fl}=\SI{-174}{\dBm}$. 

Each r-CAV randomly selected the amount of sensor data it generates from $g \in \left\lbrace \SI{0.25}{}, \SI{0.5}{}, \SI{0.75}{}, \SI{1}{} \right\rbrace$ (in \SI{}{\Gbps} per timeslot). e-CAVs constantly generated \SI{1}{\Gbps} of sensor data. The different radius used are $R \in \left\lbrace \SI{20}{\meter}, \SI{30}{\meter}, \SI{40}{\meter} \right\rbrace$. Two different beamwidths were used ($\theta=5\degree$ and $\theta=15\degree$), to investigate the effect of the different antenna beamwidths on our system. Finally, $20$ CAVs were used for our scenario. This density was carefully chosen in order to generate a small traffic jam within our network in order to exploit the increased number of links that each CAV can form.

\begin{figure*}[!htb]
\minipage{0.49\textwidth}
\centering
	\includegraphics[width=1\columnwidth]{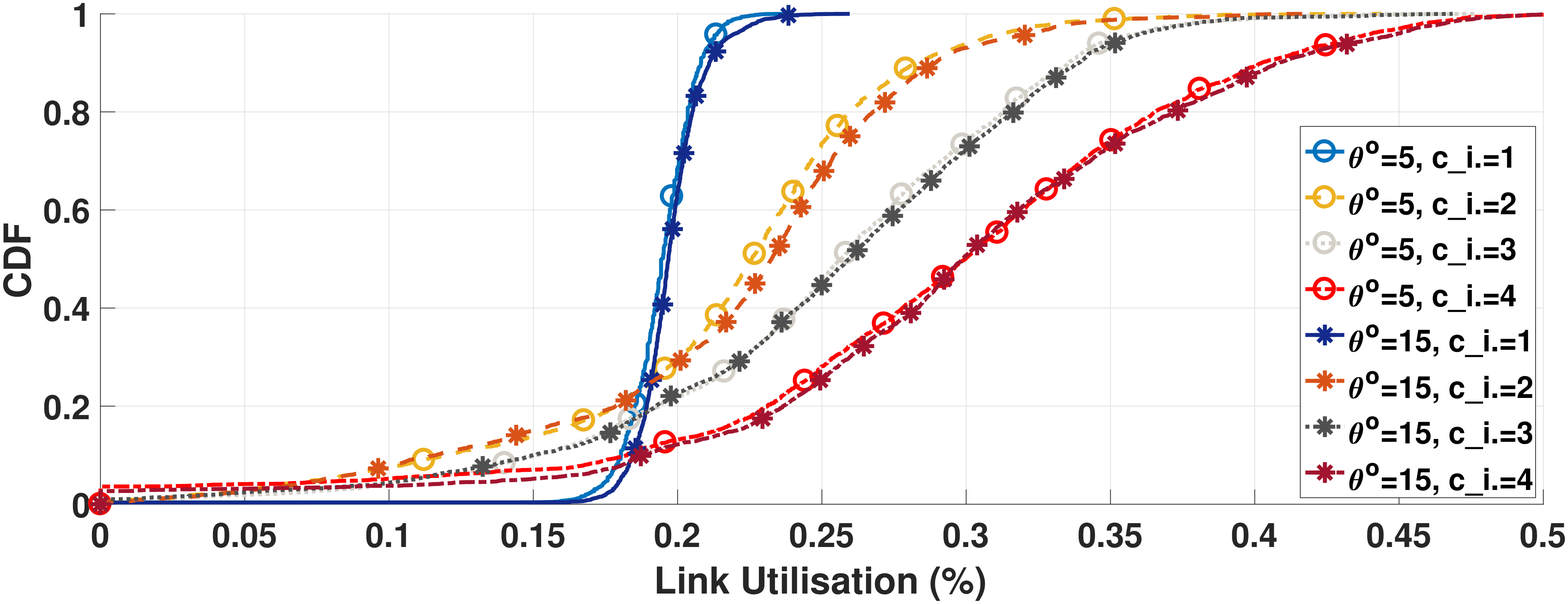}
    \vspace{-5mm}
    \caption{CDF for the link utilisation for all CAVs, and for different matching capacities and beamwidths. The radius is equal to $R=\SI{20}{\meter}$.}
    \label{fig:cdfLinkUtilRad20}
\endminipage\hfill
\minipage{0.49\textwidth}
\centering
\centering
    \includegraphics[width=1\columnwidth]{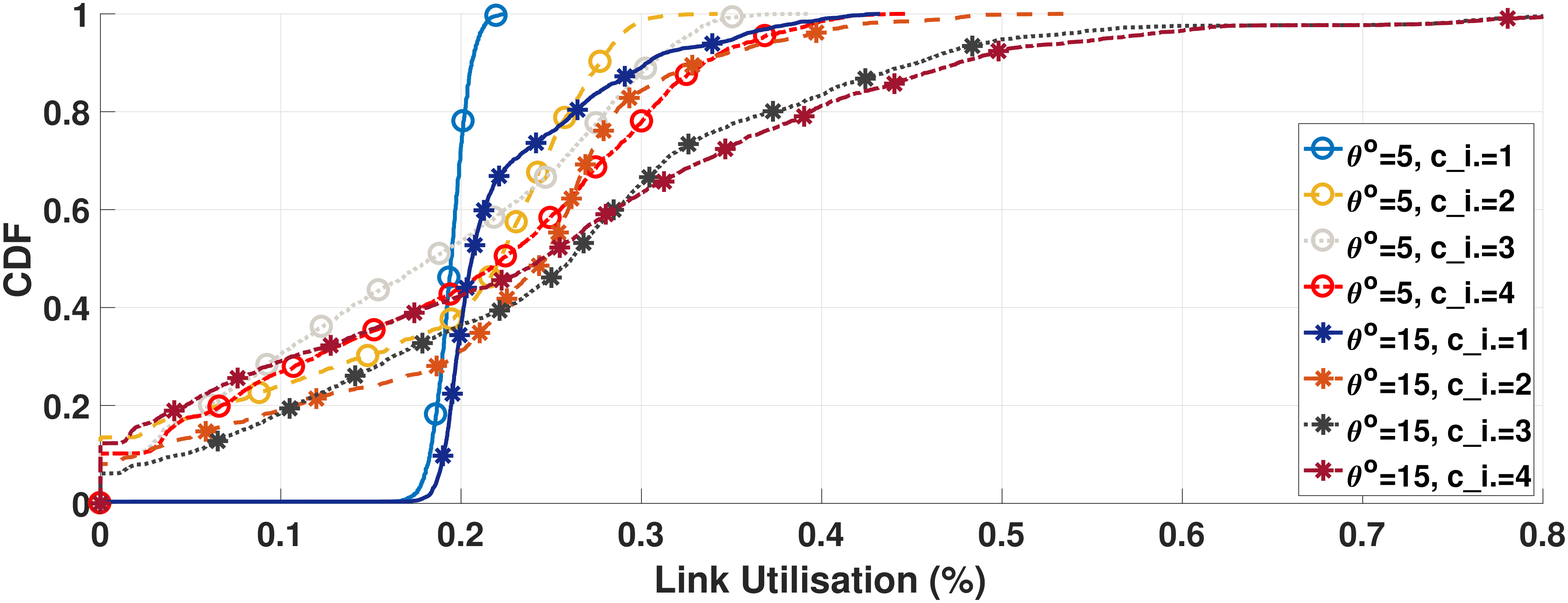}
    \vspace{-5mm}
    \caption{CDF for the link utilisation for all CAVs, and for different matching capacities and beamwidths. The radius is equal to $R=\SI{40}{\meter}$.}
    \label{fig:cdfLinkUtilRad40}
\endminipage\hfill
\minipage{0.49\textwidth}
\centering
    \includegraphics[width=1\columnwidth]{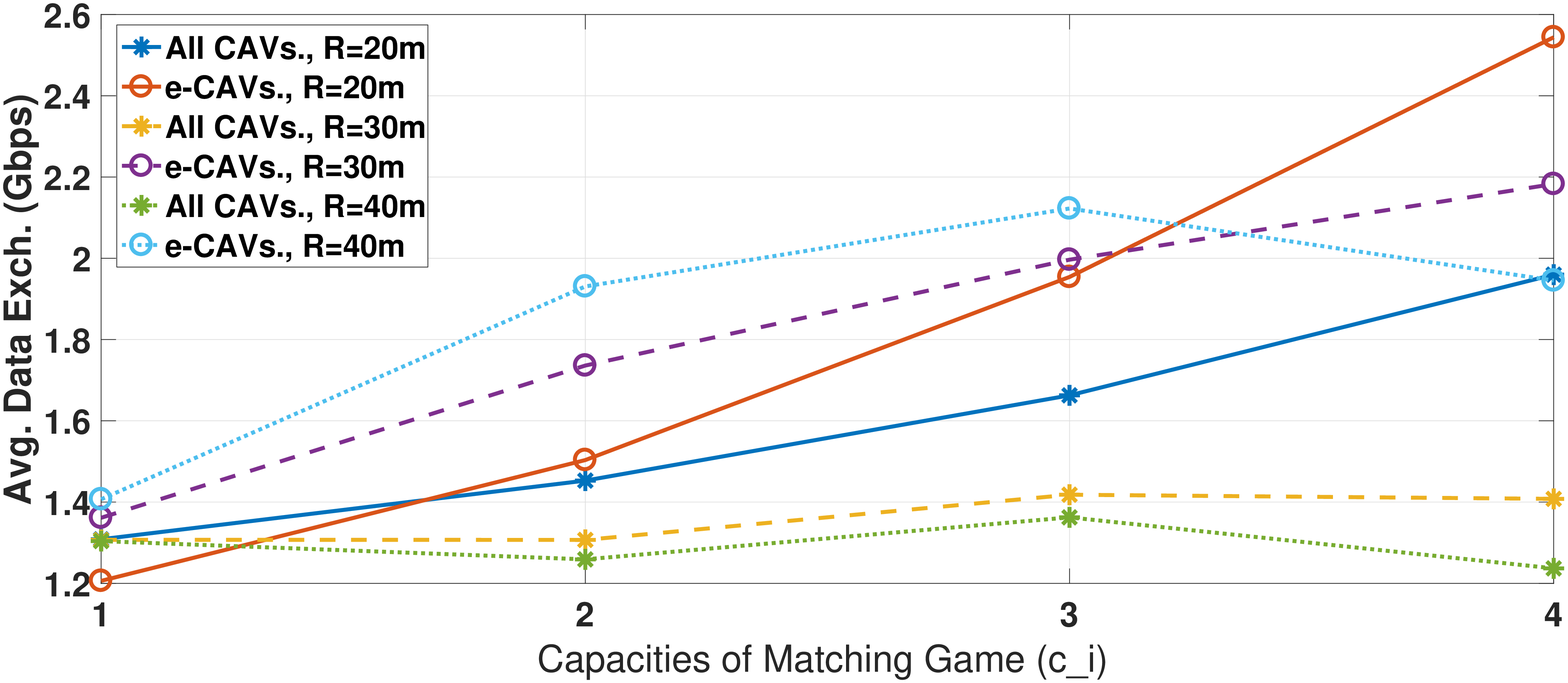}
    \vspace{-5mm}
    \caption{Average data exchanged per timeslot for each CAV (either for all CAVs or individually each e-CAV). The beamwidth is $\theta = 15\degree$.}\vspace{-3mm}
    \label{fig:avgDataTransmittedMedDens}
\endminipage\hfill
\minipage{0.49\textwidth}
\centering
  \includegraphics[width=1\columnwidth]{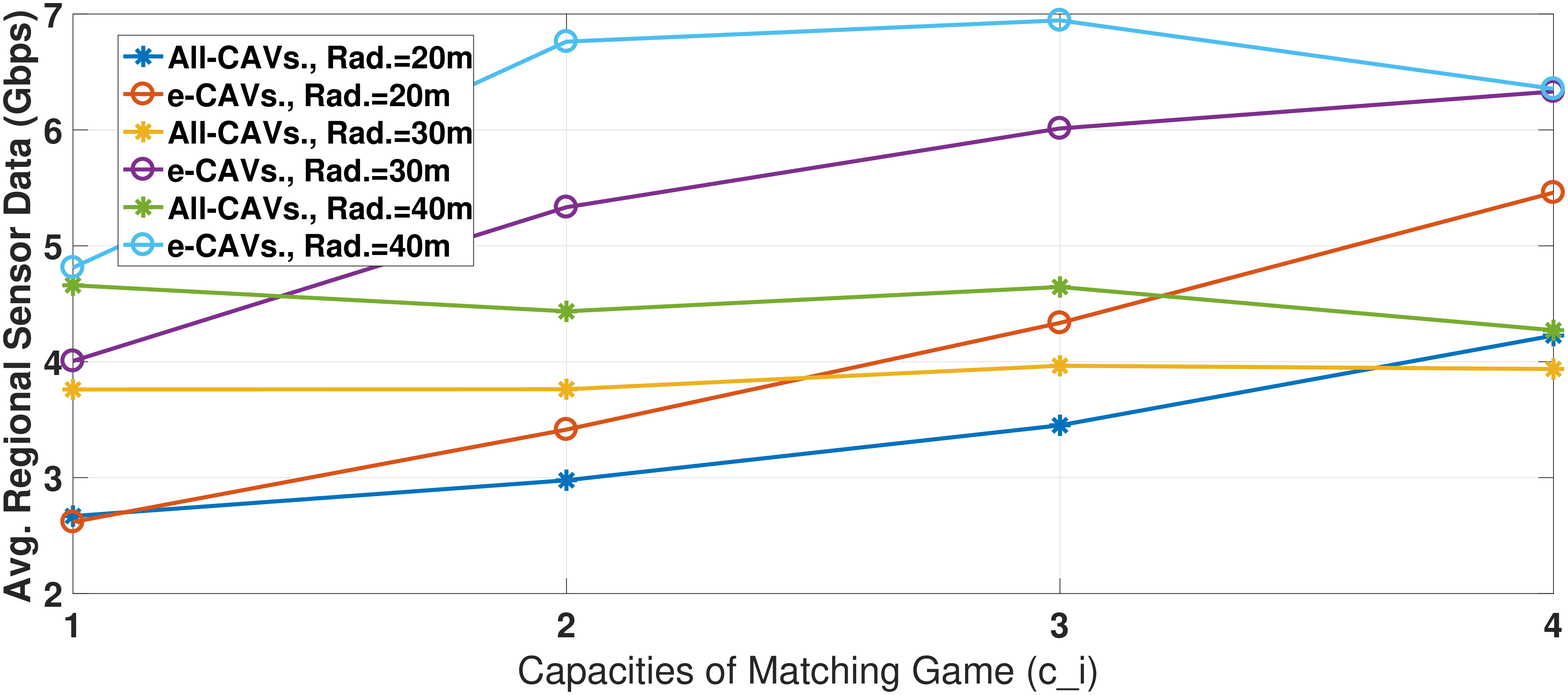}
  	\vspace{-5mm}
    \caption{Average access to sensor data per timeslot for each CAV (either for all CAVs or individually each e-CAV). The beamwidth is $\theta = 15\degree$.}\vspace{-3mm}
    \label{fig:avgRegDataMedDens}
\endminipage\hfill
\end{figure*}

\subsection{Simulation Results}\label{sub:results}

At first, we evaluated the link utilisation (per timeslot) for our system, defined as the amount of data exchanged, over $r^{\mathrm{max}}$. In Fig.~\ref{fig:cdfLinkUtilRad20}, the link utilisation for $R=\SI{20}{\meter}$ with respect to the matching game capacity and the two different beamwidths is presented. As shown, increasing the capacity, the link utilisation significantly increases. A CAV, being able to form multiple links, tends to better utilise its channel compared to the baseline case where $c_i = 1$, i.e, only point-to-point links are formed. The different beamwidths do not significantly change the results, having the case of $\theta = 15\degree$ being slightly better compared to the $\theta = 5\degree$ scenario.

In Fig.~\ref{fig:cdfLinkUtilRad40}, we present the link utilisation for a bigger radius ($R=\SI{40}{\meter}$). Again, we observe that the link utilisation increases as the matching capacity is increased. However, we notice that there is a big difference comparing the results for the two beamwidths. Using a narrow beamwidth, we increase the antenna gain and consequently our data rate. However, a greater data rate implies that this factor will dominate our utility function.
So, a CAV will prefer the surrounding vehicles providing increased data rate, i.e., the closest ones, and not the ones with the more sensor data to offer.
On the other hand, a wider beamwidth increases the link utilisation compared to the narrower case. The increased sensor information received compensates with the lower data rate due to the wider beamwidth improving our system performance.
A wider beamwidth will make our system more tolerable to beam misalignment problems.
This is very important, especially for e-CAVs where beam misalignments might lead to lost or delayed sensor information thus, delays in their maneuvers. 

Comparing Figs.~\ref{fig:cdfLinkUtilRad20} and~\ref{fig:cdfLinkUtilRad40}, it is worth noting that increasing R, the discrepancy of the link utilisation increases as well. This is because the increased number of vehicles within $R$ leads to more blocking pairs and less stable matches. For $R=\SI{30}{\meter}$, the results are similar to the above cases lying between the two aforementioned cases.


Fig.~\ref{fig:avgDataTransmittedMedDens} presents the average data exchanged (sent and received) for each CAV within a timeslot. Increasing the matching capacity compared to the baseline point-to-point links, the performance of our system is improved. What is more, comparing the exchanged data between all vehicles and just the e-CAVs individually, we observe that our approach prioritizes the traffic towards e-CAVs as intended. Starting with $c_i=1$, e-CAVs exchange roughly the same amount of data with the remaining vehicles. However, when the capacity is increased, the average exchanged data with e-CAVs becomes greater compared to the remaining vehicles. Furthermore, for $c_i=2$ and $c_i=3$, we observed better performance for $R=\SI{40}{\meter}$, whereas when $c_i=4$, the maximum achieved data exchanged was observed when $R=\SI{20}{\meter}$. This is because the increased matching capacity and radius reduce the number of stable matches, consequently changing the performance of our system. 

Similarly, in Fig.~\ref{fig:avgRegDataMedDens}, the average access to regional sensor data via the paired CAVs is presented. Again e-CAVs have access to more data compared to r-CAVs (\textasciitilde\SI{4.2}{\Gbps} to \textasciitilde\SI{5.5}{\Gbps} for $c_i=4$ and $R=\SI{20}{\meter})$. This value is almost doubled for $R=\SI{40}{\meter}$ (\textasciitilde\SI{4.6}{\Gbps} to \textasciitilde\SI{7.4}{\Gbps}). The above will have a direct impact in the manoeuvrability of e-CAVs as they will have access to more sensor data. Also, and as shown, our system does not limit the sensor data access to the remaining CAVs. The performance when $\theta=5\degree$ is similar to the aforementioned ones (Figs.~\ref{fig:avgDataTransmittedMedDens} and~\ref{fig:avgRegDataMedDens}) and due to the limited space, it will not be visually presented.

\section{Conclusions}\label{sec:conclusions}

In this work we presented an sophisticated association scheme, operating in a heterogeneous manner able to enhance the cooperative autonomous driving for the next-generation CAVs. Formulating a utility function based on the information encapsulated within DSRC beacons, we prioritized the generated sensor data towards the higher priority CAVs, without limiting the access to the rest of the network. Also, using a Stable Fixtures matching game we were able to form multipoint-to-multipoint links increasing the network throughput and the channel utilisation for all CAVs, compared to the traditional point-to-point links. The performance of our system was evaluated using realistic mobility traces. Utilising the aforementioned approach, e-CAVs at first and the remaining CAVs later, are expected to have enhanced access to valuable sensor data from their surrounding environment enhancing their cooperated maneuverability and consequently the road safety.


\vspace{-1mm}\section*{Acknowledgment}
This work was partially supported by the University of Bristol and the Engineering and Physical Sciences Research Council (EPSRC) (grant ref. EP/I028153/1). This work is also part of the FLOURISH Project, which is supported by Innovate UK under Grant 102582.

\vspace{-2mm}
\bibliographystyle{IEEEtran}
\bibliography{bib.bib,IEEEabrv}

\begin{thebibliography}{10}
\providecommand{\url}[1]{#1}
\csname url@samestyle\endcsname
\providecommand{\newblock}{\relax}
\providecommand{\bibinfo}[2]{#2}
\providecommand{\BIBentrySTDinterwordspacing}{\spaceskip=0pt\relax}
\providecommand{\BIBentryALTinterwordstretchfactor}{4}
\providecommand{\BIBentryALTinterwordspacing}{\spaceskip=\fontdimen2\font plus
\BIBentryALTinterwordstretchfactor\fontdimen3\font minus
  \fontdimen4\font\relax}
\providecommand{\BIBforeignlanguage}[2]{{%
\expandafter\ifx\csname l@#1\endcsname\relax
\typeout{** WARNING: IEEEtran.bst: No hyphenation pattern has been}%
\typeout{** loaded for the language `#1'. Using the pattern for}%
\typeout{** the default language instead.}%
\else
\language=\csname l@#1\endcsname
\fi
#2}}
\providecommand{\BIBdecl}{\relax}
\BIBdecl

\bibitem{trajectory_planning}
X.~Li, Z.~Sun, D.~Cao, Z.~He, and Q.~Zhu, ``{Real-Time Trajectory Planning for
  Autonomous Urban Driving: Framework, Algorithms, and Verifications},''
  \emph{IEEE/ASME Trans. on Mechatronics}, vol.~21, no.~2, pp. 740--753, Apr.
  2016.

\bibitem{connected_path_planning}
E.~Uhlemann, ``{Connected-Vehicles Applications Are Emerging},'' \emph{IEEE
  Veh. Technol. Mag}, vol.~11, no.~1, pp. 25--96, Mar. 2016.

\bibitem{7858628}
I.~Chatzigeorgiou and A.~Tassi, ``Decoding delay performance of random linear
  network coding for broadcast,'' \emph{IEEE Transactions on Vehicular
  Technology}, vol.~66, no.~8, pp. 7050--7060, Aug. 2017.

\bibitem{2018arXiv180109510M}
I.~{Mavromatis}, A.~{Tassi}, G.~{Rigazzi}, R.~J. {Piechocki}, and A.~{Nix},
  ``{{Multi-Radio 5G Architecture for Connected and Autonomous Vehicles:
  Application and Design Insights}},'' \emph{EAI Transactions on Industrial
  Networks and Intelligent Systems}, Jan. 2018.

\bibitem{8275608}
A.~Tassi, R.~J. Piechocki, and A.~Nix, ``{High-Speed Data Dissemination over
  Device-to-Device Millimeter-Wave Networks for Highway Vehicular
  Communication},'' in \emph{Proc. of IEEE VNC 2017}, Nov. 2017.

\bibitem{sensors}
N.~Lu, N.~Cheng, N.~Zhang, X.~Shen, and J.~W. Mark, ``{Connected Vehicles:
  Solutions and Challenges},'' \emph{IEEE Internet Things J.}, vol.~1, no.~4,
  pp. 289--299, Aug. 2014.

\bibitem{TassiTVT}
A.~Tassi, M.~Egan, R.~J. Piechocki, and A.~Nix, ``{Modeling and Design of
  Millimeter-Wave Networks for Highway Vehicular Communication},'' \emph{{IEEE}
  Trans. Veh. Technol.}, vol.~66, no.~12, Aug. 2017.

\bibitem{emergency_vehicles}
J.~Liu, P.~Jayakumar, J.~L. Stein, and T.~Ersal, ``{Combined Speed and Steering
  Control in High-Speed Autonomous Ground Vehicles for Obstacle Avoidance Using
  Model Predictive Control},'' \emph{IEEE Trans. Veh. Technol.}, vol.~66,
  no.~10, pp. 8746--8763, Oct. 2017.

\bibitem{beam_design}
V.~Va, T.~Shimizu, G.~Bansal, and R.~W. Heath, ``Beam {Design} for {Beam}
  {Switching} {Based} {Millimeter} {Wave} {Vehicle}-to-{Infrastructure}
  {Communications},'' in \emph{Proc. IEEE ICC 2016}, May 2016.

\bibitem{mywork2}
I.~Mavromatis, A.~Tassi, R.~J. Piechocki, and A.~Nix, ``{MmWave} {System} for
  {Future} {ITS}: A {MAC}-layer {Approach} for {V2X} {Beam} {Steering},'' in
  \emph{Proc. of IEEE VTC-Fall 2017}, Sep. 2017.

\bibitem{matching_theory}
Y.~Gu, W.~Saad, M.~Bennis, M.~Debbah, and Z.~Han, ``{Matching Theory for Future
  Wireless Networks: Fundamentals and Applications},'' \emph{IEEE Commun.
  Mag.}, vol.~53, no.~5, pp. 52--59, May 2015.

\bibitem{wysiwyg}
C.~Perfecto, J.~D. Ser, M.~Bennis, and M.~N. Bilbao, ``{Beyond WYSIWYG: Sharing
  Contextual Sensing Data through MmWave V2V Communications},'' in \emph{Proc.
  of EuCNC 2017}, Jun. 2017.

\bibitem{stable_fixtures}
R.~W. Irving and S.~Scott, ``{The Stable Fixtures Problem - A many-to-many
  Extension of Stable Roommates},'' \emph{Discrete Applied Mathematics}, vol.
  155, no.~16, pp. 2118 -- 2129, 2007.

\bibitem{standard}
\emph{{Enhancements} for {Very} {High} {Throughput} in the 60 {GHz} {Band}},
  Std. {IEEE} 802.11ad - 3, Mar. 2014.

\bibitem{prediction_model}
A.~Yamamoto, K.~Ogawa, T.~Horimatsu, A.~Kato, and M.~Fujise, ``Path-{Loss}
  {Prediction} {Models} for {Intervehicle} {Communication} at 60 {GHz},''
  \emph{{IEEE} Trans. Veh. Technol.}, vol.~57, no.~1, pp. 65--78, Jan. 2008.

\bibitem{balanis}
C.~A. Balanis, \emph{Antenna Theory: Analysis and Design, 4th Edition}.\hskip
  1em plus 0.5em minus 0.4em\relax John Wiley \& Sons, Mar. 2016.

\bibitem{sigma_sf}
M.~R. Akdeniz, Y.~Liu, M.~K. Samimi, S.~Sun, S.~Rangan, T.~S. Rappaport, and
  E.~Erkip, ``Millimeter wave channel modeling and cellular capacity
  evaluation,'' \emph{IEEE J. Sel. Areas Commun.}, vol.~32, no.~6, pp.
  1164--1179, Jun. 2014.

\bibitem{sumo}
D.~Krajzewicz, J.~Erdmann, M.~Behrisch, and L.~Bieker, ``{Recent Development
  and Applications of SUMO - Simulation of Urban MObility},''
  \emph{International Journal On Advances in Systems and Measurements}, vol.~5,
  no. 3\&4, pp. 128--138, Dec. 2012.

\bibitem{path_loss}
E.~Ben-Dor, T.~S. Rappaport, Y.~Qiao, and S.~J. Lauffenburger,
  ``{Millimeter}-{Wave} 60 {GHz} {Outdoor} and {Vehicle} {AOA} {Propagation}
  {Measurements} {Using} a {Broadband} {Channel} {Sounder},'' in \emph{Proc.
  IEEE GLOBECOM 2011}, Dec. 2011.

\end{thebibliography}
\end{document}